\documentclass[iop,apj,numberedappendix]{emulateapj}
\usepackage{amsmath,amssymb,color}
\usepackage{url}

\begin{document}
\title{Optical Synchrotron Precursors of Radio Hypernovae}

\author{Daisuke Nakauchi$^{1}$, Kazumi Kashiyama$^{2}$, Hiroki Nagakura$^{3}$, Yudai Suwa$^{3, 4}$ and Takashi Nakamura$^{1}$}

\altaffiltext{1}{Department of Physics, Kyoto University, Oiwake-cho, Kitashirakawa, Sakyo-ku,Kyoto 606-8502, Japan} 
\altaffiltext{2}{Einstein fellow---Department of Astronomy; Department of Physics; Theoretical Astrophysics Center; University of California, Berkeley, Berkeley, CA 94720, USA}
\altaffiltext{3}{Yukawa Institute for Theoretical Physics, Kyoto University, Oiwake-cho, Kitashirakawa, Sakyo-ku, Kyoto 606-8502, Japan}
\altaffiltext{4}{Max-Planck-Institut f\"ur Astrophysik, Karl-Schwarzschild- Str. 1, D-85748 Garching, Germany}

\begin{abstract}
We examine the bright radio synchrotron counterparts of low-luminosity gamma-ray bursts~(llGRBs) and relativistic supernovae~(SNe) and find that they can be powered by spherical hypernova~(HN) explosions.
Our results imply that radio-bright HNe are driven by relativistic jets that are choked deep inside the progenitor stars or quasi-spherical magnetized winds from fast-rotating magnetars.
We also consider the optical synchrotron counterparts of radio-bright HNe and show that they can be observed as precursors several days before the SN peak with an $r$-band absolute magnitude of $M_r \sim -14$ mag.
While previous studies suggested that additional trans-relativistic components are required to power the bright radio emission, we find that they overestimated the energy budget of the trans-relativistic component by overlooking some factors related to the minimum energy of non-thermal electrons.
If an additional trans-relativistic component exists, then a much brighter optical precursor with $M_r \sim -20$ mag can be expected.
Thus, the scenarios of radio-bright HNe can be distinguished by using optical precursors, which can be detectable from $\lesssim 100\ \rm Mpc$ by current SN surveys like the Kiso SN Survey, Palomar Transient Factory, and Panoramic Survey Telescope \& Rapid Response System.
\end{abstract}

\maketitle

\section{Introduction}
A good fraction of core-collapse supernovae~(SNe) has bright radio counterparts called radio SNe.
The radio emission is  due to the synchrotron emission from non-thermal electrons accelerated at the shock 
with a velocity of $v \sim 0.1 c$~\citep[e.g.,][]{Chevalier1982, Chevalier1998}.
Such fast-moving ejecta are  formed when SN shocks break out of the progenitor stars~\citep{Matzner1999, Tan2001}. 
Therefore, radio SNe are  good probes of the dynamics of the SN ejecta, the progenitor structure, and  the circumstellar medium~\citep[CSM; e.g.,][]{Weiler2002}.

Intrinsically much brighter radio counterparts have been observed in several broad-lined Type Ibc SNe~(SNe Ibc) or hypernovae~(HNe).
Previous authors claimed that these radio emissions are too bright to be powered by the ejecta produced by SN/HN shock breakout 
so that  additional trans-relativistic components are required~\citep{Soderberg2010, Chakraborti2011, Chakraborti2014, Margutti2014, Milisavljevic2015}. 
They proposed that relativistic jets which barely punch out the progenitor stars are  the origins of the trans-relativistic components~\citep{Margutti2014}.
In fact, some of these radio-bright HNe are associated with low-luminosity gamma-ray bursts~(llGRBs), while others, like SN 2009bb and SN 2012ap, did not show detectable high-energy emission.
The latter events are called relativistic SNe. 

\begin{figure}[!t]
\begin{center}
\includegraphics[scale=0.3, angle = -90]{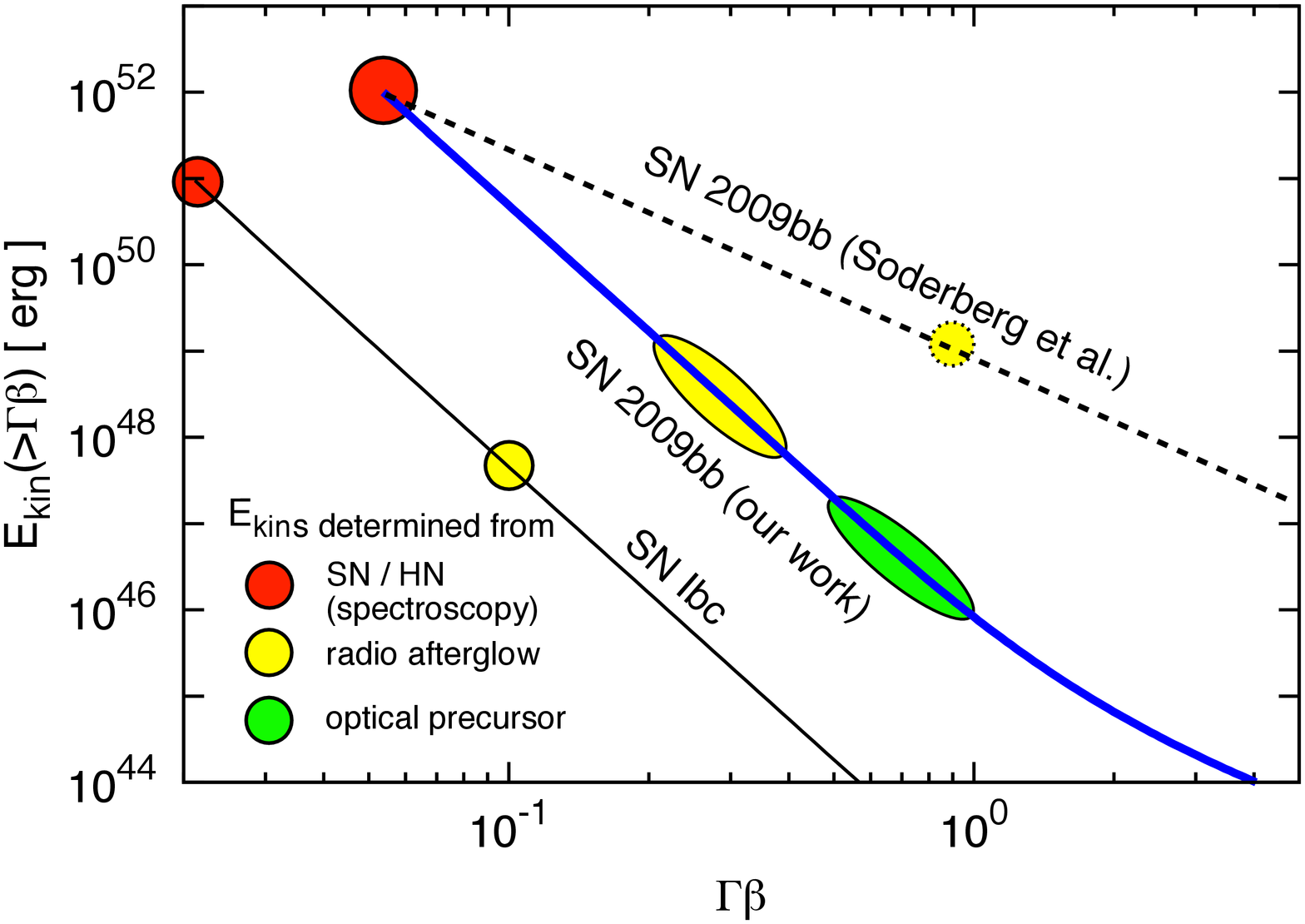}
\caption{Energy profile of a SN/HN ejecta as a function of $\Gamma \beta$.
The red points show the total energy of an SN/HN $E_{\rm in}$, and the solid lines show the profiles theoretically predicted from spherical SN/HN explosions~(Equations \eqref{eq:ene_dist}-\eqref{eq:e_tilde}).
The black line corresponds to the representative case of an SN Ibc with $(E_{\rm in}, M_{\rm ej}) \sim (10^{51}\ {\rm erg}, 3\ M_\odot)$, 
and the blue line to an HN with $(E_{\rm in}, M_{\rm ej}) \sim (10^{52}\ {\rm erg}, 4.8\ M_\odot)$, which is consistent with SN 2009bb.
The yellow point on the dashed line corresponds to the energy of the trans-relativistic ejecta which is estimated by \cite{Soderberg2010}.
The yellow and green regions on the blue line show the shells contributing to the radio~(at 10-1000 days) and optical~(at 0.01-1 days) synchrotron emissions, respectively. They are determined on the basis of the refreshed shock model in this paper.
}
\label{Fig:ene_dist}
\end{center}
\end{figure}

To clarify the above arguments, in Figure \ref{Fig:ene_dist}, we show the energy profile of SN/HN ejecta as a function of $\Gamma \beta$, where $\beta = v/c$, and $\Gamma = 1/\sqrt{1-\beta^2}$ is the Lorentz factor.
The solid lines correspond to the energy profiles of a normal SN Ibc~(black) and an HN~(SN 2009bb; blue), which are theoretically predicted from spherical explosions~\citep{Matzner1999, Tan2001}.
On the other hand, the yellow point on the dashed line was obtained by \cite{Soderberg2010} to explain the bright radio counterpart of SN 2009bb.  
As one can see, there is a significant gap between the blue line and the yellow point on the dashed line. 
This is the reason why the previous authors introduced an additional trans-relativistic component  driven by a relativistic jet~(the dashed line).
If this is the case, then radio-bright HNe may be a missing link between ordinary SNe Ibc and HNe associated with GRBs.

In this paper, however, we show that the previous studies overestimated the energy and the speed of the trans-relativistic ejecta.
In Section \ref{sec:RSM}, we describe the refreshed shock model of a spherical SN/HN ejecta. This model is used to calculate the emission from a radio-bright HN in Section \ref{sec:Result}.
In Section \ref{subsec:radio}, we first estimate the energy profile of HN ejecta, on the basis of the refreshed shock model.
We find that the energy  profile is consistent with that predicted from the spherical HN explosion~(the blue solid line in Figure \ref{Fig:ene_dist}).
Then, we point out  that the previous authors overlooked some factors related to the minimum energy of the non-thermal electrons.
In Section \ref{subsec:optical}, we consider the optical counterpart of a radio-bright HN, and show that it can be observed at $0.01\mbox{-}1\ \rm days$ after the shock breakout as the precursor of SN emission.
Such optical precursors can be detected using current and future SN surveys and provide further insight into the explosion mechanism of HNe and the circumstellar environments.
In particular, the detection of an optical precursor can be crucial to distinguish between our model and the previous one. 
In Section \ref{subsec:parameter}, we discuss the effect of the phenomenological parameters on our results.
Section \ref{sec:Discussion} is devoted to the summary and discussion.

\section{Model}\label{sec:RSM}
\subsection{Dynamics}\label{sec:dynamics}
First, we model the energy profile of the ejecta produced by a spherical SN/HN explosion~(Section \ref{sec:profile}), and then we model the deceleration of such ejecta in the CSM~(Section \ref{sec:decelerate}).

\subsubsection{Ejecta Profile}\label{sec:profile}
Let us consider an SN/HN explosion with a total energy of $E_{\rm in}$ and an ejecta mass of $M_{\rm ej}$.
The SN/HN blast wave is accelerated as it propagates through the outer envelope of the progenitor where the density declines steeply~\citep{Sakurai1960, Johnson1971}.
A small fraction of the surface layer can be accelerated up to trans-relativistic velocities, $\Gamma \beta \sim 1$. 
After the breakout, the shocked ejecta are further accelerated by converting the internal energy into the kinetic energy.
The resultant cumulative kinetic-energy distribution can be described as~(\citealt{Matzner1999,Tan2001}; the solid lines in Figure 1)\footnote{
$\Tilde{E}$ and $F(\Gamma \beta)$ also depend on the progenitor structure, for which we adopt the same stripped-envelope progenitor as in \cite{Tan2001}.
Following their convention, we assume the following set of parameters: $q = 4.1$, $\gamma_{\rm p} = 4/3$, $C_{\rm nr} = 2.03$, $f_{\rho} = 0.63$, $f_{\rm sph} = 0.85$ and $A = 0.736$.}
\begin{equation}
E_{\rm kin}( > \Gamma \beta) = \Tilde{E}  F(\Gamma \beta),
\label{eq:ene_dist}
\end{equation}
where $F(\Gamma \beta)$  is a decreasing function of $\Gamma \beta$ and is given in Equation (38) of \cite{Tan2001} as\footnote{
More precisely, the proportionality coefficient of Equation \eqref{eq:ene_dist2} is not a constant, but a complex function of $\Gamma \beta$~\citep{Tan2001}. 
This evaluation is valid for $\Gamma \beta \lesssim 1$.
}
\begin{equation}
F(\Gamma \beta) \sim 20\ [(\Gamma \beta)^{-3.85/4.1} + (\Gamma \beta)^{-0.83/4.1}]^{16.4/3},
\label{eq:ene_dist2}
\end{equation}
and the energy coefficient, $\Tilde{E}$, is evaluated as 
\begin{equation}
\Tilde{E} \sim 5.5 \times 10^{40}\ \left(\frac{E_{\rm in}}{10^{51}\ {\rm erg}}\right)^{\frac{10.7}{3}} \left(\frac{M_{\rm ej}}{3\ M_\odot}\right)^{-\frac{7.7}{3}}\ \rm erg.
\label{eq:e_tilde}
\end{equation}
In Figure \ref{Fig:ene_dist}, we show the representative case of an SN Ibc~(the black line) with $(E_{\rm in}, M_{\rm ej}) = (10^{51}\ {\rm erg}, 3\ M_\odot)$, and an HN~(the blue line) with $(E_{\rm in}, M_{\rm ej}) = (10^{52}\ {\rm erg}, 4.8\ M_\odot)$, which is consistent with SN 2009bb~\citep{Pignata2011}. 

\subsubsection{Dynamics of the Decelerating Ejecta}\label{sec:decelerate}
We assume a power law for the CSM density profile, $n_{\rm w}(R) = A_2 R^{-2}$, 
where $R$ is the radius and $A_2 = \dot{M}/(4 \pi v_{\rm w} m_{\rm p})$ with $\dot{M}$, $v_{\rm w}$, and $m_{\rm p}$ being the mass loss rate, 
the wind velocity, and the proton mass, respectively. 
Here, we fix the wind velocity as $v_{\rm w} = 1000\ {\rm km}\ {\rm s}^{-1}$, which is a typical value for Wolf-Rayet~(W-R) stars~\citep[e.g.,][]{Crowther2007}.

Since the outer shells have larger velocities and smaller energies, they decelerate first by interacting with the CSM.
The decelerated shells constitute a shocked region.  
The inner, slower shells successively catch up with the shocked region and energize it~\citep[refreshed shock;][]{Rees1998}.
The total energy in the shocked region can be calculated as~\citep{Taylor1950, Sedov1959, Blandford1976, De_Colle2012}
\begin{align}
E_{\rm sh}(\Gamma \beta, R) &= R^{3} (\Gamma \beta)^2 n_{\rm w}(R) m_{\rm p} c^2 \notag\\ 
& \times \left[\frac{8 \pi}{9} \beta^2 + \frac{9}{4 \alpha_2} (1-\beta^2)\right], 
\label{eq:E_int}
\end{align}
for a shock velocity $\beta c$  and radius $R$. 
Here, $\alpha_2^{1/3} = 0.78$.\footnote{Equation \eqref{eq:E_int} reproduces the numerical results of blast wave evolution 
within a maximum difference of 5\%, both in the trans-relativistic and non-relativistic regimes~\citep{De_Colle2012}.}
As long as radiative cooling is negligible, $E_{\rm sh}(\Gamma \beta, R)$ is equal to the original kinetic energy $E_{\rm kin}(> \Gamma \beta)$, 
so that the shock velocity can be estimated from~\citep{Sari2000, Kyutoku2014, Duran2015}
\begin{equation}
E_{\rm kin}(> \Gamma \beta) = E_{\rm sh}(\Gamma \beta, R).
\label{eq:vel_rad}
\end{equation}
By integrating $dR/dt = \beta(R)c$ with respect to the lab-frame time $t$, 
the shock radius can be obtained  as a function of $t$ as $R = R(t)$.
Moreover, the lab-frame time can be related to the observer-frame time $t_{\rm obs}$ through $dt_{\rm obs}/dt = 1-\beta(R)$.  

In the non-relativistic limit, Equation \eqref{eq:vel_rad} can be approximately represented as
\begin{equation}
20 \tilde{E} \beta^{1-s_{\rm nr}} \sim 9 R^{3} \beta^2 n_{\rm w}(R) m_{\rm p} c^2 / 4 \alpha_2,
\label{eq:vel_rad_non_rela}
\end{equation}
where $s_{\rm nr} = 18.4/3$.
From this, the time evolution of the blast wave radius can be calculated from
\begin{align}
R(t_{\rm obs}) & \sim [(4 \alpha_2/9c)(20 \tilde{E}/A_2 m_{\rm p} c^2)]^{1/(s_{\rm nr}+2)} \notag \\
& \times c\ (t_{\rm obs}/\tilde{s}_{\rm nr})^{\tilde{s}_{\rm nr}},
\label{eq:radius}
\end{align}
where $\tilde{s}_{\rm nr} = (s_{\rm nr}+1)/(s_{\rm nr}+2)$.
After the shock velocity becomes smaller than that of the slowest shell, the evolution can be described by the Sedov-von Neumann-Taylor solutions with an energy of $E_{\rm in}$.

\subsection{Synchrotron Emission}\label{subsec:emission}
Next, we model the synchrotron emission from the decelerating SN/HN ejecta based on the external shock model of non-relativistic fireballs~\citep{Waxman1998, Frail2000, Sironi2013}. 
We note that while the equations below explicitly contain the Lorentz factor $\Gamma$ of the shocked fluid, the non-relativistic regime can be consistently calculated by approximating $\Gamma \approx 1$ and $\Gamma - 1 \approx \beta^2/2$.

We make a few assumptions for simplicity~\citep{Sironi2013}.
First, we take as constants the fractions of the internal energy in the post-shock fluid that are used to generate turbulent magnetic fields and to accelerate non-thermal electrons, $\epsilon_B$ and $\epsilon_e$, respectively.
Second, all of the electrons that are swept up by the blast wave are accelerated.
Third, the injection spectrum of the non-thermal electrons is a single power law with an index of $p$: 
$n(\gamma_e) d \gamma_e = n_0 \gamma_e^{-p} d \gamma_e$ ($\gamma_e \geq \gamma_{\rm m}$), 
where $\gamma_e$ is the Lorentz factor of the non-thermal electrons, $\gamma_{\rm m}$ is the minimum value,  
$n(\gamma_e)$ is the number density of the accelerated electrons, and $n_0$ is the normalization factor. 

The number density and the internal energy density of the post-shock fluid are calculated from~\citep{Blandford1976}
\begin{equation}
n_{\rm ps} = 4 \Gamma n_{\rm w},
\label{eq:n_ps}
\end{equation}
and
\begin{equation}
e_{\rm int} = 4 \Gamma (\Gamma-1) n_{\rm w} m_{\rm p} c^2,
\label{eq:e_int}
\end{equation}
respectively.
From the assumptions above, $\int_{\gamma_{\rm m}}^{\infty} n_0 \gamma_e^{-p} d \gamma_e = 4 \Gamma n_{\rm w}$ 
and $\int_{\gamma_{\rm m}}^{\infty} n_0 \gamma_e^{1-p} m_e c^2 d \gamma_e = \epsilon_e e_{\rm int}$ where $m_e$ is the electron mass. 
Then, $n_0$ and $\gamma_{\rm m}$ can be calculated from
\begin{equation}
n_0 = (p-1) \gamma_{\rm m}^{p-1} 4 \Gamma n_{\rm w},
\label{eq:n_0}
\end{equation}
and
\begin{equation}
\gamma_{\rm m} = 1 + \frac{m_{\rm p}}{m_{\rm e}} \frac{p-2}{p-1} \epsilon_e (\Gamma-1).
\label{eq:gamma_m}
\end{equation}
The magnetic field strength is calculated from $B^2 / 8 \pi = \epsilon_B e_{\rm int}$ as
\begin{equation}
B = [8 \pi m_{\rm p} c^2 \epsilon_B n_{\rm w} 4 \Gamma (\Gamma-1)]^{1/2}.
\label{eq:mag}
\end{equation}
The electron energy spectrum becomes a broken power law due to significant synchrotron cooling above a critical Lorentz factor $\gamma_{\rm c}$, 
where an electron loses almost all of the energy within the dynamical time:  
\begin{equation}
\gamma_{\rm c} = \frac{6 \pi m_{\rm e} c}{\sigma_T B^2 \Gamma t }. 
\label{eq:gamma_c}
\end{equation}
The synchrotron frequencies corresponding to electrons with $\gamma_{m}$ and $\gamma_{c}$ are  
\begin{equation}
\nu_{\rm m, c} = \nu(\gamma_{\rm m, c}) = \frac{QB}{2 \pi m_{\rm e} c} \gamma_{\rm m, c}^2 \Gamma,
\label{eq:nu_mc}
\end{equation}
where $Q$ is the elemental charge.
Synchrotron self-absorption~(SSA) becomes important at radio frequencies.
The optical depth for SSA can be calculated from $\tau(\nu) \sim \alpha(\nu) R / \Gamma$~\citep{Panaitescu2000, Inoue2004}, 
where $\alpha(\nu)$ is the absorption coefficient~\citep{Rybicki1979}. 
Note that the width of the shocked region can be evaluated from $\Delta R \sim R/\Gamma^2$ in the lab frame, and $\Delta R' \sim R / \Gamma$ in the comoving frame.
We determine the absorption frequency $\nu_{\rm a}$ from $\tau(\nu_{\rm a}) = 1$.

We consider the emission only from the forward shock region.
In the observer frame, the total synchrotron power emitted from a relativistic electron with $\gamma_e$ is given by $P(\gamma_e) = (4\sigma_{\rm T} c/3)(B^2/8\pi) \gamma_e^2 \Gamma^2$~\citep{Rybicki1979}.
Since the emitted photon energy concentrates around the typical synchrotron frequency $\nu(\gamma_e)$, the spectral peak power from a single electron $P_{\nu, {\rm max}}$ can be calculated from
\begin{equation}
P_{\nu, {\rm max}} \sim \frac{P(\gamma_e)}{\nu(\gamma_e)} = \frac{m_e c^2 \sigma_{\rm T}}{3 Q} B \Gamma.
\label{eq:peak_power}
\end{equation}
If self-absorption is negligible, then the peak flux density $F_{\nu, {\rm max}}$ from all of the non-thermal electrons can be calculated from
\begin{equation}
F_{\nu, {\rm max}} \sim \frac{P_{\nu, {\rm max}} 4 \pi n_{\rm w} R^3}{4 \pi D^2}, 
\label{eq:peak_flux}
\end{equation}
where $D$ is the distance to the source and $\int_0^R 4 \pi R'^2 n_{\rm w}(R') dR' = 4 \pi n_{\rm w} R^3$ is the total number of the swept-up electrons.

In the synchrotron emission model, the spectral energy distribution~(SED) has three break frequencies, $\nu_{\rm a}$, $\nu_{\rm m}$, and $\nu_{\rm c}$~\citep{Sari1998, Panaitescu2000, Inoue2004}.
In the radio-emitting phase, the following inequality holds among these frequencies: $\nu_{\rm m} < \nu_{\rm a} < \nu_{\rm c}$.
In this case, the SED can be approximately calculated from the broken power law~\citep{Sari1998, Granot2002}
\begin{equation}
F_{\nu} \sim F_{\nu, {\rm max}} \begin{cases}
\left(\frac{\nu_{\rm a}}{\nu_{\rm m}} \right)^{-(p-1)/2}\left(\frac{\nu}{\nu_{\rm a}} \right)^{5/2}, & \nu_{\rm m} < \nu < \nu_{\rm a}, \\
\left(\frac{\nu}{\nu_{\rm m}} \right)^{-(p-1)/2}, & \nu_{\rm a} < \nu < \nu_{\rm m}, \\
\left(\frac{\nu_{\rm c}}{\nu_{\rm m}} \right)^{-(p-1)/2}\left(\frac{\nu}{\nu_{\rm c}} \right)^{-p/2}, & \nu_{\rm c} < \nu. \\
\end{cases}
\label{eq:spectrum}
\end{equation}
The SED peaks at the absorption frequency
\begin{equation}
\nu_{\rm p} = \nu_{\rm a},
\label{eq:peak_freq_abs}
\end{equation}
where the peak flux density $F_{\rm p}$ can be calculated from
\begin{equation}
F_{\rm p} = F_{\nu_{\rm a}} \sim F_{\nu, {\rm max}}(\nu_{\rm a}/\nu_{\rm m})^{-(p-1)/2}.
\label{eq:peak_flux_abs}
\end{equation}

Our model has five input parameters: $\Tilde{E}$, $\epsilon_e$, $\epsilon_B$, $p$, and $\dot{M}$~(or $A_2$).
On the other hand, the observed radio spectrum is essentially characterized by three parameters, $\nu_{\rm p}$, $F_{\rm p}$, 
and the spectral slope, which can be associated with $\nu_{\rm a}$, $F_{\nu_{\rm a}}$, and $p$, through Equations \eqref{eq:spectrum}-\eqref{eq:peak_flux_abs}.
Thus, from radio observations, one can determine $\Tilde{E}$ and $\dot{M}$~(or $A_2$) for a given set of $\epsilon_e$ and $\epsilon_B$~(see Equations \ref{eq:E_tilde} and \ref{eq:Mdot_tilde} for the explicit forms).

\section{Result}\label{sec:Result}
First, in Section \ref{subsec:radio}, we first estimate the energy profile of the ejecta of a radio-bright HN from the observed radio spectrum, on the basis of the refreshed shock model in the previous section.
We focus on SN 2009bb at $D = 40$ Mpc, since the observed data are available from \cite{Soderberg2010}. 
We find that the energy profile is consistent with that predicted from the spherical HN explosion.
Then, we discuss the origin of the difference between our results and the previous studies.
In Section \ref{subsec:optical}, we focus on the synchrotron emission at optical frequencies and suggest that it can be an important counterpart for discriminating the origin of radio-bright HNe.
Finally, in Section \ref{subsec:parameter}, we discuss the impact of the phenomenological parameters on our results.

\subsection{Radio Afterglow}\label{subsec:radio}
\begin{figure}[!t]
\begin{center}
\includegraphics[scale=1.0]{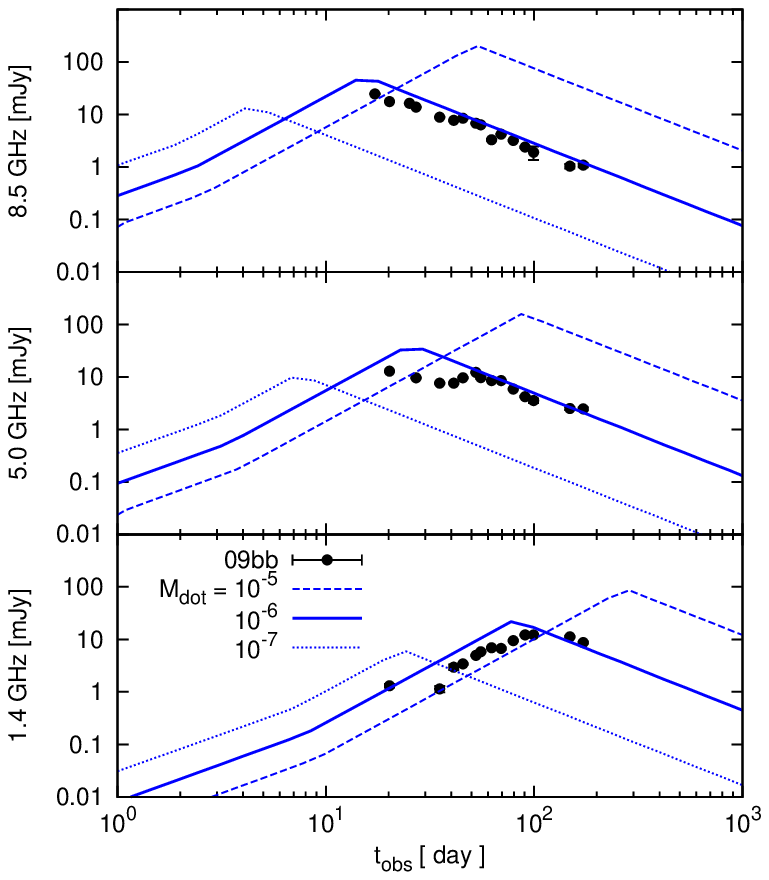}
\caption{Radio light curve of SN 2009bb.  
The black data points are taken from \cite{Soderberg2010}.
The solid lines show our theoretical fit with $\epsilon_e = \epsilon_B = 0.33$, $p = 3$, $\Tilde{E} = 6 \times 10^{43}\ {\rm erg}$, and $\dot{M} = 10^{-6}\ M_{\odot}\ {\rm yr}^{-1}$.  
Note that $\Tilde{E} = 6 \times 10^{43}\ {\rm erg}$ corresponds to the blue solid line in Figure \ref{Fig:ene_dist}. 
The dashed and dotted lines show a higher~($\dot{M} = 10^{-5}\ M_{\odot}\ {\rm yr}^{-1}$) and a lower~($\dot{M} = 10^{-7}\ M_{\odot}\ {\rm yr}^{-1}$) mass loss cases, respectively.}
\label{Fig:LC_radio}
\end{center}
\end{figure}

Here, to begin, we estimate the energy profile of a HN ejecta from the observed radio spectrum on the basis of the refreshed shock model in the previous section.
By substituting Equations \eqref{eq:radius}-\eqref{eq:peak_flux} into Equations \eqref{eq:peak_freq_abs} and \eqref{eq:peak_flux_abs}, we can estimate $\tilde{E}$ and $\dot{M}$ as functions of $\epsilon_e$ and $\epsilon_B$ as
\begin{align}
\tilde{E} &\sim 6 \times 10^{43}\ \left[\left(\frac{\epsilon_B}{0.33}\right)^{(s_{\rm nr}+10)(1-p)} \left(\frac{\epsilon_e}{0.33}\right)^{(s_{\rm nr}-4p+1)}\right]^{1/(4p+9)} \notag \\ 
& \times \left(\frac{\nu_{\rm p}}{6\ {\rm GHz}} \frac{t_{\rm fit}}{20\ {\rm day}}\right)^{(-s_{\rm nr}(2p+13)+24p-31)/(4p+9)} \notag \\
& \times \left(\frac{F_{\rm p}}{20\ {\rm mJy}}\right)^{(s_{\rm nr}(p+6)-6(p-4))/(4p+9)}\ \rm erg \notag \\
& \propto \epsilon_B^{-14.6/63} \epsilon_e^{-96.8/63},
\label{eq:E_tilde}
\end{align}
and
\begin{align}
\dot{M} & \sim 10^{-6}\ \left(\frac{\epsilon_B}{0.33}\right)^{-(4p+1)/(4p+9)} \left(\frac{\epsilon_e}{0.33}\right)^{-8(p-1)/(4p+9)} \notag \\ 
& \times \left(\frac{\nu_{\rm p}}{6\ {\rm GHz}} \frac{t_{\rm fit}}{20\ {\rm day}}\right)^{2(12p-7)/(4p+9)} \notag \\
& \times \left(\frac{F_{\rm p}}{20\ {\rm mJy}}\right)^{-4(2p-3)/(4p+9)}\ M_\odot\ {\rm yr}^{-1} \notag \\
&\propto \epsilon_B^{-13/21} \epsilon_e^{-16/21},
\label{eq:Mdot_tilde}
\end{align}
where $F_{\rm p} \sim 20\ {\rm mJy}$ and $\nu_{\rm p} \sim 6\ {\rm GHz}$ are the peak flux density and the peak frequency determined from the radio spectrum at $t_{\rm fit} \sim 20$ days in \cite{Soderberg2010}.
Note that  $s_{\rm nr} = 18.4/3$ from Equation \eqref{eq:vel_rad_non_rela} and the spectral index is obtained as $p \sim 3$ from the observed spectral slope~\citep{Soderberg2010}.
On the other hand, the plausible values of the phenomenological parameters $\epsilon_{e}$ and $\epsilon_B$, are uncertain.
Here, we adopt the equipartition values, since the main aim is to compare our estimate of the energy profile with that of \cite{Soderberg2010}.
We discuss the impact of the phenomenological parameters on our results in Section \ref{subsec:parameter}.
Note that the energy profile with $\Tilde{E} = 6 \times 10^{43}\ {\rm erg}$ corresponds to the blue solid line in Figure \ref{Fig:ene_dist}.
Therefore, our results imply that the energy profile of a radio-bright HN, SN 2009bb, is consistent with that predicted from the spherical HN explosion, and it does not require the additional trans-relativistic component.

In Figure \ref{Fig:LC_radio}, we compare the light curves of SN 2009bb in the radio band with those calculated by the refreshed shock model.  
The black points correspond to the observed data from~\cite{Soderberg2010}.
The solid blue lines are the results of our theoretical calculation with $\Tilde{E} = 6 \times 10^{43}\ {\rm erg}$, $\dot{M} = 10^{-6}\ M_{\odot}\ {\rm yr}^{-1}$, $\epsilon_e = \epsilon_B = 0.33$, and $p = 3$.
For comparison, we also show the cases of the higher~($\dot{M} = 10^{-5}\ M_{\odot}\ {\rm yr}^{-1}$) and lower~($\dot{M} = 10^{-7}\ M_{\odot}\ {\rm yr}^{-1}$) mass loss rates with the dashed and dotted lines, respectively.
The radio flux becomes larger and the peak time comes later for the higher wind density.
One can see that the radio counterpart of SN 2009bb can be well explained by the refreshed shock model using the estimated energy profile, CSM density, and the adopted equipartition parameters. 

Next, let us compare the obtained energy profile with that of the previous studies.
Using the yellow and green regions on the blue solid line of Figure \ref{Fig:ene_dist}, we show the shells contributing to the radio and optical synchrotron emission, respectively.
The radio-emitting shells have $\Gamma \beta \sim 0.4\mbox{-}0.2$ and the cumulative energies of $E_{\rm sh} \sim 10^{48}\mbox{-}10^{49}\ {\rm erg}$ for $t_{\rm obs} \sim 10\mbox{-}10^3$ days~(the yellow region on the solid line).
On the other hand, \cite{Soderberg2010} estimated $\Gamma \beta \sim 0.85$, $E_{\rm sh} \sim 10^{49}\ {\rm erg}$~(the yellow point on the dashed line), and $\dot{M} = 2 \times 10^{-6}\ M_{\odot}\ {\rm yr}^{-1}$ by fitting the radio spectrum of SN 2009bb at $t_{\rm obs} \sim 20$ days.
We find, however, that they may overestimate $E_{\rm sh}$ and $\Gamma \beta$ by overlooking some factors related to the minimum Lorentz factor~($\gamma_{\rm m}$) of the non-thermal electrons.
Hereafter, we discuss the origin of the discrepancy in their estimate following their arguments.

By replacing $E_l$~(the minimum energy of the non-thermal electrons) in Equations (11) and (12) of \cite{Chevalier1998} with $\gamma_{\rm m, fit} m_e c^2$, we obtain the emission radius $R_{\rm fit}$ and the magnetic field strength  $B_{\rm fit}$ as
\begin{align}
R_{\rm fit} &\sim 3.8 \times 10^{16}\ \left(\frac{F_{\rm p}}{20\ {\rm mJy}}\right)^{9/19} \left(\frac{\nu_{\rm p}}{6\ {\rm GHz}}\right)^{-1} \notag \\ 
& \times \left(\frac{\epsilon_B}{0.33}\right)^{1/19} \left(\frac{\epsilon_e}{0.33}\right)^{-1/19} \gamma_{\rm m, fit}^{-1/19}\ \rm cm,
\label{eq:R_fit}
\end{align}
and
\begin{align}
B_{\rm fit} &\sim 0.48\ \left(\frac{F_{\rm p}}{20\ {\rm mJy}}\right)^{-2/19} \left(\frac{\nu_{\rm p}}{6\ {\rm GHz}}\right) \notag \\ 
& \times \left(\frac{\epsilon_B}{0.33}\right)^{4/19} \left(\frac{\epsilon_e}{0.33}\right)^{-4/19} \gamma_{\rm m, fit}^{-4/19}\ \rm G,
\label{eq:B_fit}
\end{align}
respectively.
We can see that $R_{\rm fit}$ and  $B_{\rm fit}$ at $t_{\rm fit}$ weakly depend on $\gamma_{\rm m, fit} \equiv \gamma_{\rm m}(t_{\rm fit})$.
This is also pointed out in \cite{Chevalier2006}.
From Equation \eqref{eq:R_fit}, $\Gamma \beta$ at $t_{\rm fit}$ can be estimated as
\begin{align}
(\Gamma \beta)_{\rm fit} &\sim R_{\rm fit} / c t_{\rm fit} \notag \\
&\sim 0.73\ \left(\frac{F_{\rm p}}{20\ {\rm mJy}}\right)^{9/19} \left(\frac{\nu_{\rm p}}{6\ {\rm GHz}} \frac{t_{\rm fit}}{20\ {\rm day}} \right)^{-1} \notag \\ 
& \times \left(\frac{\epsilon_B}{0.33}\right)^{1/19} \left(\frac{\epsilon_e}{0.33}\right)^{-1/19} \gamma_{\rm m, fit}^{-1/19}.
\label{eq:vel_fit}
\end{align}
The blast wave energy is given by $E_{\rm sh} \sim R_{\rm fit}^3 B_{\rm fit}^2 / 12 \epsilon_B$, and substituting Equations \eqref{eq:R_fit} and \eqref{eq:B_fit}, it is evaluated as
\begin{align}
E_{\rm sh} &\sim 3.1 \times 10^{48}\ \left(\frac{F_{\rm p}}{20\ {\rm mJy}}\right)^{23/19} \left(\frac{\nu_{\rm p}}{6\ {\rm GHz}}\right)^{-1} \notag \\ 
& \times \left(\frac{\epsilon_B}{0.33}\right)^{-8/19} \left(\frac{\epsilon_e}{0.33}\right)^{-11/19} \gamma_{\rm m, fit}^{-11/19}\ \rm erg.
\label{eq:Esh_fit}
\end{align}
Finally, from the definition of the mass loss rate $\dot{M} = 4 \pi R^2 \rho_{\rm w} v_{\rm w}$ and Equation \eqref{eq:mag}, 
one can obtain  $\dot{M} = (v_{\rm w}/8 \epsilon_B c^2) (B^2 R^2 / \Gamma(\Gamma-1)) \sim (v_{\rm w} / 4 \epsilon_B) t_{\rm obs}^2 B^2$, or 
\begin{align}
\dot{M} & \sim 1.9 \times 10^{-6}\ \left(\frac{F_{\rm p}}{20\ {\rm mJy}}\right)^{-4/19} \left(\frac{\nu_{\rm p}}{6\ {\rm GHz}}\frac{t_{\rm fit}}{20\ {\rm day}}\right)^2 \notag \\ 
& \times \left(\frac{\epsilon_B}{0.33}\right)^{-11/19} \left(\frac{\epsilon_e}{0.33}\right)^{-8/19} \gamma_{\rm m, fit}^{-8/19}\ M_\odot\ {\rm yr}^{-1},
\label{eq:Mdot_fit}
\end{align}
where we use $R \sim \beta c t_{\rm obs}$ in the non-relativistic limit. 
Equations \eqref{eq:Esh_fit} and \eqref{eq:Mdot_fit} show that the blast wave energy~($E_{\rm sh}$) and the CSM density~($\dot{M}$) strongly depend on $\gamma_{\rm m, fit}$.

If we set $\gamma_{\rm m, fit} = 1$, then we can reproduce the estimates of \cite{Soderberg2010} by a factor of less than a few from Equations \eqref{eq:vel_fit}-\eqref{eq:Mdot_fit}.
According to Equation \eqref{eq:gamma_m}, however, we should set $\gamma_{\rm m, fit} \sim 100$ for $p = 3$, $\epsilon_e = \epsilon_B = 0.33$, and $\Gamma \beta = 0.85$, which they adopted in their study.
If we substitute $\gamma_{\rm m, fit} \sim 100$ into Equations \eqref{eq:vel_fit}-\eqref{eq:Mdot_fit}, then we obtain $\Gamma \beta \sim 0.57$, $E_{\rm sh} \sim 2.2 \times 10^{47}\ {\rm erg}$, and $\dot{M} = 2.7 \times 10^{-7}\ M_{\odot}\ {\rm yr}^{-1}$.
Thus, \cite{Soderberg2010}  overestimated $\Gamma \beta$, $E_{\rm sh}$, and  $\dot{M}$ by overlooking the large factor related to $\gamma_{\rm m, fit}$.
If they correct this point, their results are consistent with ours. 

\subsection{Optical Synchrotron Precursor}\label{subsec:optical}
\begin{figure}[!t]
\begin{center}
\includegraphics[scale=1.0]{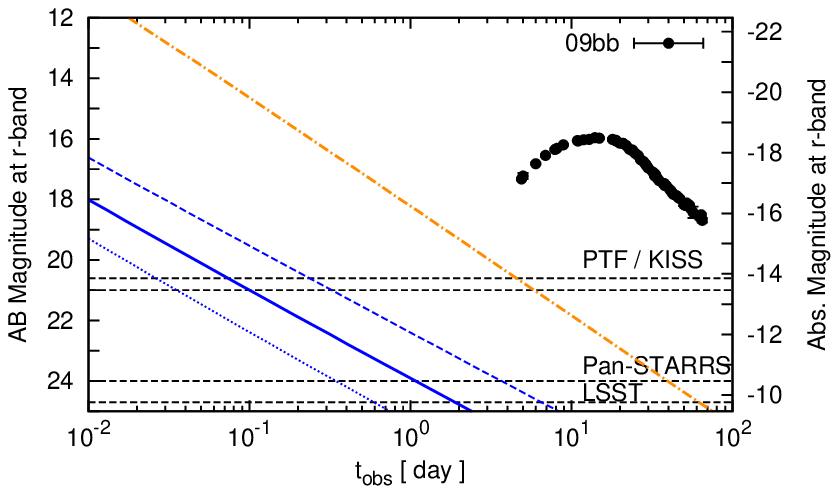}
\caption{Optical synchrotron precursor expected from the radio observation of SN 2009bb~(the blue solid line). 
The black points correspond to the $r$-band light curve of SN 2009bb~\citep{Pignata2011} and the dashed lines to the 5$\sigma$ sensitivity of 
PTF~(60 s), KISS~(180 s), Pan-STARRS~(30 s), and LSST~(30 s) from up to bottom, respectively, where the values in the parentheses correspond to the integration times.
We see that an optical synchrotron precursor is predicted against the canonical SN emission for $t_{\rm obs} < 1$ day. 
Especially for $t_{\rm obs} \lesssim 0.1$ day, such a precursor may be detected by the current detectors.  
For comparison, we also show the results of higher~($\dot{M} = 10^{-5}\ M_{\odot}\ {\rm yr}^{-1}$) and lower~($\dot{M} = 10^{-7}\ M_{\odot}\ {\rm yr}^{-1}$) mass loss cases with the dashed and dotted lines, respectively.
The optical synchrotron precursor will evolve as the orange dotted-dashed line if the estimates of \cite{Soderberg2010} were correct.
Note that in Section \ref{subsec:radio}, we show that they overestimated the energy of the HN ejecta.
Future observations of SN 2009bb-like events can confirm whether our predictions or theirs are correct.}
\label{Fig:LC_opt}
\end{center}
\end{figure}

We can see from the blue line in Figure \ref{Fig:ene_dist} that the trans-relativistic ejecta with $\Gamma \beta \sim 1$ still have a large amount of energy.
Emission from such trans-relativistic ejecta can be expected from earlier times at higher frequencies compared to the radio emission.
Here, we consider the synchrotron emission at optical frequencies.
Since a frequency in the optical band $\nu_{\rm opt}$ is found to be larger than $\nu_{\rm a}$, $\nu_{\rm m}$, and $\nu_{\rm c}$ at all times, the light curve can be calculated from the equation given by the last column of Equation \eqref{eq:spectrum}, i.e., $\nu_{\rm c} < \nu_{\rm opt}$.

In Figure \ref{Fig:LC_opt}, the solid blue line represents the optical synchrotron flux calculated from the above parameter values, 
and the black points represent the $r$-band light curve of SN 2009bb~\citep{Pignata2011}.
Here, we adopt the color excess of $E_{B-V} = 0.58$~\citep{Pignata2011}.
The dashed lines show the 5$\sigma$ sensitivity of the Palomar Transient Factory~\citep[PTF, 60 s;][]{Law2009, Rau2009}, 
Kiso Supernova Survey~\citep[KISS, 180 s;][]{Morokuma2014}, Panoramic Survey Telescope \& Rapid Response System~\citep[Pan-STARRS, 30 s;][]{Kaiser2002}, 
and Large Synoptic Survey Telescope~(LSST, 30 s)\footnote{\url{http://www.lsst.org/lsst/}} from top to bottom, respectively, 
where the values in the parentheses correspond to the integration times.
We find that at $\sim$ 0.01-1 days after shock breakout, such optical synchrotron emission can be seen as precursors of canonical HN emission.  
Especially for $t_{\rm obs} \lesssim 0.1$ day, such precursors can be detectable even using current detectors.
For comparison, we also show the cases of higher~($\dot{M} = 10^{-5}\ M_{\odot}\ {\rm yr}^{-1}$) and lower~($\dot{M} = 10^{-7}\ M_{\odot}\ {\rm yr}^{-1}$) mass loss rates with the dashed and dotted lines, respectively.
Brighter precursors can be expected for the denser wind envelopes, and so it can be a good probe of the circumstellar environments.
Note that the shells contributing to the optical precursor have velocities larger than those contributing to the radio afterglow: the former have $\Gamma \beta \sim 1\mbox{-}0.5$ and the cumulative energies of $E_{\rm sh} \sim 10^{46}\mbox{-}10^{47}\ {\rm erg}$ for $t_{\rm obs} \sim$ 0.01-1 days~(the green region in Figure \ref{Fig:ene_dist}).

An optical precursor can also be expected from the energy profile of \cite{Soderberg2010}, which is shown to be overestimated in Section \ref{subsec:radio}.
In Figure \ref{Fig:LC_opt}, the orange dotted-dashed line is the optical precursor calculated from their energy profile.
We can see that their optical precursor is by $\sim$ 6 AB magnitude brighter than ours, so that we will be able to distinguish between our estimate and theirs from the future observations of the SN 2009bb-like events.
More generally, one can test whether a radio-bright HN really requires an additional trans-relativistic component when we combine the observations of an optical precursor, SN emission, and a radio afterglow. Thus, the detection of an optical precursor can be crucial to determine the explosion mechanism of a radio-bright HN.

\subsection{Dependence on Phenomenological Parameters}\label{subsec:parameter}
So far, we adopt the equipartition values for the phenomenological parameters, $\epsilon_e = \epsilon_B = 0.33$, 
since our main goal is to compare our estimate of the energy profile with that of \cite{Soderberg2010}.
The plausible values of these parameters are, however, rather uncertain.
For example, from the combined analysis of the late-time radio and X-ray emission of Type IIb SNe, 
lower values of $\epsilon_e$ are obtained, $\epsilon_e \sim 0.01, \epsilon_B \sim 0.1$~\citep{Maeda2012}, 
while GRB afterglows show the opposite trends: $\epsilon_e \sim 0.1, \epsilon_B \sim 0.01$~\citep{Panaitescu2002, Yost2003}.

We can estimate the larger values of $\tilde{E}$ and $\dot{M}$ for the smaller values of $\epsilon_e$ and $\epsilon_B$~(see Equations \ref{eq:E_tilde} and \ref{eq:Mdot_tilde}).
For example, for $\epsilon_e \sim 0.01, \epsilon_B \sim 10^{-3}$, we can obtain $\dot{M} \sim 5.2 \times 10^{-4}\ M_\odot\ {\rm yr}^{-1}$ and $\tilde{E} \sim 5 \times 10^{46}\ \rm erg$.
In this case, the resultant energy profile passes through the yellow point in Figure \ref{Fig:ene_dist}, 
while the wind mass loss rate is a bit larger than those of the Galactic W-R stars: $\dot{M}_{\rm WR} < 10^{-4}\ M_\odot\ {\rm yr}^{-1}$~\citep{Crowther2007, Smith2014}.
If this is the case, then we can suggest that a radio-bright HN may require an additional trans-relativistic component.

We also confirm that while the brightness of the optical precursor tends to become dimmer for the smaller values of $\epsilon_e$ and $\epsilon_B$, 
the difference is at most of 1 AB magnitude, and that it is still be detectable even by the current detectors at $t_{\rm obs} \gtrsim$ 0.01 day after shock breakout.

\section{Discussion}\label{sec:Discussion}
Previous studies claimed that radio-bright HNe cannot be powered by the ejecta produced by a spherical HN explosion, and that an additional trans-relativistic component is required.
They proposed that relativistic jets that barely punch out the progenitor stars can be the origin of the trans-relativistic component.
In this paper, however, we focus on a radio-bright HN and find that they  overestimated the energy and the speed of the trans-relativistic ejecta, since they overlooked some factors related to $\gamma_{\rm m, fit}$.  
In addition to the radio afterglow, we also consider the optical counterpart of a radio-bright HN and find that it can be observed as the precursor of canonical SN emission by current and future SN surveys.
An optical precursor can also be expected from the energy profile of the previous studies.
We find that if their estimates were correct, then we would see an optical precursor that is by $\sim$ 6 AB magnitude brighter than ours.
Therefore, the detection of an optical precursor can be crucial to distinguish between our estimate and the previous one.
More generally, one can test whether a radio-bright HN really has an additional trans-relativistic component by combining the observations of an optical precursor, SN emission, and a radio afterglow.

We find that even the current SN surveys can detect 09bb-like optical precursors at the very early times up to $\sim 100$ Mpc.
Since the fraction of 09bb-like HNe is $\sim 0.7$ \% of SNe Ibc~\citep{Soderberg2010}, we may expect a good event rate of $\lesssim 0.5\ {\rm yr}^{-1}$ for 09bb-like optical counterparts from PTF and KISS.
In the LSST era, we can expect the detection of optical precursors not only from more distant events but also from ordinary HNe.
They would provide deeper insight into the GRB-SN connection.
Note that they would not be hidden by the SN shock breakout emission since its duration and spectrum peak are expected to be $R_{\rm WR} / c \sim 1\mbox{-}10$ s and to be in the UV to X-ray bands, if we consider a typical stripped-envelope WR progenitor~\citep{Chevalier2008}.

Recently, \cite{Duran2015} calculated the afterglow emissions of relativistic shock breakout based on the refreshed shock model.
They showed that both the prompt and afterglow emission of llGRBs can be consistently explained in the framework of relativistic shock breakout~\citep{Nakar2012}. 
However, they mainly focused on the llGRBs and late-time~($\gtrsim 1$ day) afterglow emissions, and did not consider the trans-relativistic motion, which is relevant to our study.

Our results and those of \cite{Duran2015} imply that the central engine of radio-bright HNe should produce quasi-spherical ejecta. One possible scenario for the central engine is that the relativistic jet is choked deep within the progenitor star, since the duration of the central engine activity is much shorter than the breakout timescale, as discussed in \cite{Lazzati2012}.
Another possibility is that the central engine is a rapidly rotating magnetar that generates a quasi-spherical outflow~\citep[e.g.,][]{Thompson2004}.

At the early evolution stage~($t_{\rm obs} \lesssim 0.3$ day), when the blast wave radius is still small and the CSM is dense enough, 
we find that the absorption frequency $\nu_{\rm a}$ becomes larger than the cooling frequency $\nu_{\rm c}$.  
In this case, self-absorption may become a heating source for the accelerated electrons.
Electrons are piled up at a Lorentz factor where self-absorption heating and synchrotron cooling balance each other~\citep{McCray1969, Ghisellini1988}.
Moreover, the radiation spectrum approaches a quasi-thermal spectrum for $\nu < \nu_{\rm a}$.
We find, however, that the absorption frequency may always be smaller than any frequency $\nu_{\rm opt}$ in the optical band: $\nu_{\rm opt} > \nu_{\rm a} \sim 2 \times 10^{12} (t_{\rm obs}/0.1\ {\rm day})^{-0.84}\ {\rm Hz}$ for fiducial parameters $\epsilon_e = \epsilon_B = 0.33$, 
$\Tilde{E} = 6 \times 10^{43}\ {\rm erg}$, and $\dot{M} = 10^{-6}\ M_{\odot}\ {\rm yr}^{-1}$.  
Therefore, self-absorption heating may not qualitatively change the optical precursor in Figure \ref{Fig:LC_opt}.

We also check that inverse Compton~(IC) emission does not significantly vary our results as long as we adopt the fiducial parameters of equipartition.
Here, we consider the synchrotron-self-Compton~(SSC) emission and external-IC~(EIC) emission.
We find that SSC emission is weak for all of the time, since the Compton $Y$ parameter can be evaluated as $Y_{\rm SSC} < 0.62$~\citep{Sari2001}.
For EIC emission, since SN thermal photons dominate the external radiation field, we should compare the energy density of SN thermal photons $U_{\rm rad} \sim 0.11 (L_{\rm bol}/10^{42.7}\ {\rm erg}\ {\rm s}^{-1})(t_{\rm obs}/10\ {\rm day})^{-1.8}\ {\rm erg}\ {\rm cm}^{-3}$ with that of the magnetic filed $U_{B} \sim 0.034\ (t_{\rm obs}/10\ {\rm day})^{-2}\ {\rm erg}\ {\rm cm}^{-3}$, where $L_{\rm bol}$ is the bolometric peak luminosity.
As we can see from Figure \ref{Fig:LC_opt}, for $t_{\rm obs} > 50$ days, the SN becomes so dim that $U_{\rm rad} \ll U_{B}$, and EIC emission can be negligible, while for $t_{\rm obs} \lesssim 50$ days, the SN is still bright enough that $U_{\rm rad} \gtrsim U_{B}$, and EIC can be the dominant cooling process. 
Since the observed radio light curve can be reproduced quite well for $t_{\rm obs} > 50$ days with the synchrotron emission model~(Figure \ref{Fig:LC_radio}),
EIC does not affect for determining the model parameters.

\section*{Acknowledgements}
We thank the anonymous referee for giving us helpful comments and improving the quality of this paper.
We also thank K. Kyutoku and K. Maeda for fruitful discussions and
comments.  This work is supported in part by the Grant-in-aid from the
Ministry of Education, Culture, Sports, Science and Technology (MEXT)
of Japan, Nos. 261051 (DN), 24740165, 24244036 (HN), 25103511 (YS), 23540305, 24103006 (TN), HPCI Strategic Program of MEXT (HN), NASA through Einstein Postdoctoral Fellowship grant number PF4-150123 awarded by the Chandra X-ray Center, which is operated by the Smithsonian Astrophysical Observatory for NASA under contract NAS8-03060 (KK), JSPS postdoctoral fellowships for research abroad, MEXT SPIRE, and JICFuS.


\begin{thebibliography}{99}

\bibitem[Barniol Duran et al.(2015)]{Duran2015} Barniol Duran, 
R., Nakar, E., Piran, T., \& Sari, R.\ 2015, \mnras, 448, 417

\bibitem[Blandford 
\& McKee(1976)]{Blandford1976} Blandford, R.~D., \& McKee, C.~F.\ 1976, Physics of Fluids, 19, 1130

\bibitem[Chakraborti 
\& Ray(2011)]{Chakraborti2011} Chakraborti, S., \& Ray, A.\ 2011, \apj, 729, 57 

\bibitem[Chakraborti et al.(2014)]{Chakraborti2014} Chakraborti, S., 
Soderberg, A., Chomiuk, L., et al.\ 2014, arXiv:1402.6336 

\bibitem[Chevalier(1982)]{Chevalier1982} Chevalier, R.~A.\ 1982, 
\apj, 258, 790

\bibitem[Chevalier(1998)]{Chevalier1998} Chevalier, R.~A.\ 1998, 
\apj, 499, 810

\bibitem[Chevalier 
\& Fransson(2006)]{Chevalier2006} Chevalier, R.~A., \& Fransson, C.\ 2006, \apj, 651, 381

\bibitem[Chevalier 
\& Fransson(2008)]{Chevalier2008} Chevalier, R.~A., \& Fransson, C.\ 2008, \apjl, 683, L135

\bibitem[Crowther(2007)]{Crowther2007} Crowther, P.~A.\ 2007, \araa, 45, 177

\bibitem[Frail et al.(2000)]{Frail2000} Frail, D.~A., Waxman, E., 
\& Kulkarni, S.~R.\ 2000, \apj, 537, 191

\bibitem[De Colle et al.(2012)]{De_Colle2012} De Colle, F., Granot, 
J., L{\'o}pez-C{\'a}mara, D., \& Ramirez-Ruiz, E.\ 2012, \apj, 746, 122

\bibitem[Ghisellini et al.(1988)]{Ghisellini1988} Ghisellini, G., 
Guilbert, P.~W., \& Svensson, R.\ 1988, \apjl, 334, L5

\bibitem[Granot \& Sari(2002)]{Granot2002} Granot, J., \& Sari, R.\ 2002, \apj, 568, 820

\bibitem[Inoue(2004)]{Inoue2004} Inoue, S.\ 2004, \mnras, 348, 999

\bibitem[Johnson 
\& McKee(1971)]{Johnson1971} Johnson, M.~H., \& McKee, C.~F.\ 1971, \prd, 3, 858

\bibitem[Kaiser et al.(2002)]{Kaiser2002} Kaiser, N., Aussel, H., 
Burke, B.~E., et al.\ 2002, \procspie, 4836, 154

\bibitem[Kyutoku et al.(2014)]{Kyutoku2014} Kyutoku, K., Ioka, K., 
\& Shibata, M.\ 2014, \mnras, 437, L6

\bibitem[Law et al.(2009)]{Law2009} Law, N.~M., Kulkarni, 
S.~R., Dekany, R.~G., et al.\ 2009, \pasp, 121, 1395

\bibitem[Lazzati et al.(2012)]{Lazzati2012} Lazzati, D., Morsony, 
B.~J., Blackwell, C.~H., \& Begelman, M.~C.\ 2012, \apj, 750, 68

\bibitem[Maeda(2012)]{Maeda2012} Maeda, K.\ 2012, \apj, 758, 81

\bibitem[Margutti et al.(2014)]{Margutti2014} Margutti, R., 
Milisavljevic, D., Soderberg, A.~M., et al.\ 2014, \apj, 797, 107

\bibitem[Matzner 
\& McKee(1999)]{Matzner1999} Matzner, C.~D., \& McKee, C.~F.\ 1999, \apj, 510, 379

\bibitem[McCray(1969)]{McCray1969} McCray, R.\ 1969, \apj, 156, 
329

\bibitem[Milisavljevic et al.(2015)]{Milisavljevic2015} Milisavljevic, 
D., Margutti, R., Parrent, J.~T., et al.\ 2015, \apj, 799, 51

\bibitem[Morokuma et al.(2014)]{Morokuma2014} Morokuma, T., 
Tominaga, N., Tanaka, M., et al.\ 2014, arXiv:1409.1308

\bibitem[Nakar 
\& Sari(2012)]{Nakar2012} Nakar, E., \& Sari, R.\ 2012, \apj, 747, 88

\bibitem[Panaitescu 
\& Kumar(2000)]{Panaitescu2000} Panaitescu, A., \& Kumar, P.\ 2000, \apj, 543, 66

\bibitem[Panaitescu 
\& Kumar(2002)]{Panaitescu2002} Panaitescu, A., \& Kumar, P.\ 2002, \apj, 571, 779

\bibitem[Pignata et al.(2011)]{Pignata2011} Pignata, G., 
Stritzinger, M., Soderberg, A., et al.\ 2011, \apj, 728, 14

\bibitem[Rau et al.(2009)]{Rau2009} Rau, A., Kulkarni, S.~R., 
Law, N.~M., et al.\ 2009, \pasp, 121, 1334

\bibitem[Rees 
\& M{\'e}sz{\'a}ros(1998)]{Rees1998} Rees, M.~J., \& Meszaros, P.\ 1998, \apjl, 496, L1

\bibitem[Rybicki 
\& Lightman(1979)]{Rybicki1979} Rybicki, G.~B., \& Lightman, A.~P.\ 1979, New York, Wiley-Interscience, 1979.~393 p.,  

\bibitem[Sakurai(1960)]{Sakurai1960} Sakurai A., 1960, Commun. Pure Appl. Math., 13, 353

\bibitem[Sari 
\& Esin(2001)]{Sari2001} Sari, R., \& Esin, A.~A.\ 2001, \apj, 548, 787

\bibitem[Sari et al.(1998)]{Sari1998} Sari, R., Piran, T., 
\& Narayan, R.\ 1998, \apjl, 497, L17 

\bibitem[Sari 
\& M{\'e}sz{\'a}ros(2000)]{Sari2000} Sari, R., \& M{\'e}sz{\'a}ros, P.\ 2000, \apjl, 535, L33

\bibitem[Sedov(1959)]{Sedov1959} Sedov, L.~I.\ 1959, Similarity 
and Dimensional Methods in Mechanics, New York: Academic Press, 1959

\bibitem[Sironi 
\& Giannios(2013)]{Sironi2013} Sironi, L., \& Giannios, D.\ 2013, \apj, 778, 107

\bibitem[Smith(2014)]{Smith2014} Smith, N.\ 2014, \araa, 52, 487

\bibitem[Soderberg et al.(2010)]{Soderberg2010} Soderberg, A.~M., 
Chakraborti, S., Pignata, G., et al.\ 2010, \nat, 463, 513 

\bibitem[Tan et al.(2001)]{Tan2001} Tan, J.~C., Matzner, C.~D., 
\& McKee, C.~F.\ 2001, \apj, 551, 946

\bibitem[Taylor(1950)]{Taylor1950} Taylor, G.\ 1950, Royal Society 
of London Proceedings Series A, 201, 159

\bibitem[Thompson et al.(2004)]{Thompson2004} Thompson, T.~A., 
Chang, P., \& Quataert, E.\ 2004, \apj, 611, 380

\bibitem[Waxman et al.(1998)]{Waxman1998} Waxman, E., Kulkarni, 
S.~R., \& Frail, D.~A.\ 1998, \apj, 497, 288

\bibitem[Weiler et 
al.(2002)]{Weiler2002} Weiler, K.~W., Panagia, N., Montes, M.~J., \& Sramek, R.~A.\ 2002, \araa, 40, 387

\bibitem[Yost et al.(2003)]{Yost2003} Yost, S.~A., Harrison, 
F.~A., Sari, R., \& Frail, D.~A.\ 2003, \apj, 597, 459

\end{thebibliography}
\end{document}